\pdfoutput=1

\documentclass[reprint,aps,prl,twocolumn,,floatfix,superscriptaddress]{revtex4-1}
\usepackage{graphicx}
\usepackage{dcolumn}
\usepackage{bm}
\usepackage{amsmath}
\usepackage{lineno}
\usepackage[english]{babel}
\usepackage[utf8]{inputenc}
\usepackage{color}

\begin{document}

\title{Effect of quantum Hall edge strips on valley splitting in silicon quantum wells}

\author{Brian Paquelet Wuetz}
\affiliation{QuTech and Kavli Institute of Nanoscience, Delft University of Technology, PO Box 5046, 2600 GA Delft, The Netherlands}
\author{Merritt P. Losert}
\affiliation{University of Wisconsin-Madison, Madison, WI 53706 USA}
\author{Alberto Tosato}
\affiliation{QuTech and Kavli Institute of Nanoscience, Delft University of Technology, PO Box 5046, 2600 GA Delft, The Netherlands}
\author{Mario Lodari}
\affiliation{QuTech and Kavli Institute of Nanoscience, Delft University of Technology, PO Box 5046, 2600 GA Delft, The Netherlands}
\author{Peter L. Bavdaz}
\affiliation{QuTech and Kavli Institute of Nanoscience, Delft University of Technology, PO Box 5046, 2600 GA Delft, The Netherlands}
\author{Lucas Stehouwer}
\affiliation{QuTech and Kavli Institute of Nanoscience, Delft University of Technology, PO Box 5046, 2600 GA Delft, The Netherlands}
\author{Payam Amin} 
\affiliation{Components Research, Intel Corporation, 2501 NW 229th Ave, Hillsboro, OR 97124, USA}
\author{James S. Clarke} 
\affiliation{Components Research, Intel Corporation, 2501 NW 229th Ave, Hillsboro, OR 97124, USA}
\author{Susan N. Coppersmith}
\affiliation{University of New South Wales, Sydney, Australia\\}
\author{Amir Sammak}
\affiliation{QuTech and Netherlands Organisation for Applied Scientific Research (TNO), Stieltjesweg 1, 2628 CK Delft, The Netherlands}
\author{Menno Veldhorst}
\affiliation{QuTech and Kavli Institute of Nanoscience, Delft University of Technology, PO Box 5046, 2600 GA Delft, The Netherlands}
\author{Mark Friesen}
\affiliation{University of Wisconsin-Madison, Madison, WI 53706 USA}
\author{Giordano Scappucci}
\email{g.scappucci@tudelft.nl}
\affiliation{QuTech and Kavli Institute of Nanoscience, Delft University of Technology, PO Box 5046, 2600 GA Delft, The Netherlands}

\date{\today}

\begin{abstract}
We determine the energy splitting of the conduction-band valleys in two-dimensional electrons confined to low-disorder Si quantum wells. We probe the valley splitting dependence on both perpendicular magnetic field $B$ and Hall density by performing activation energy measurements in the quantum Hall regime over a large range of filling factors. The mobility gap of the valley-split levels increases linearly with $B$ and is strikingly independent of Hall density. The data are consistent with a transport model in which valley splitting depends on the incremental changes in density $eB/h$ across quantum Hall edge strips, rather than the bulk density. Based on these results, we estimate that the valley splitting increases with density at a rate of 116 $\mu$eV/10$^{11}$cm$^{-2}$, consistent with theoretical predictions for near-perfect quantum well top interfaces.
\end{abstract}

\maketitle
Silicon has proven to be a successful material platform for obtaining high-fidelity electron spin-qubits in quantum dots\cite{veldhorst_addressable_2014,Yoneda2018A99.9,yang_silicon_2019}. The advanced level of quantum control in these qubits makes it possible to execute two-qubit logic gates and rudimentary quantum algorithms\cite{veldhorst_two-qubit_2015,Watson2018ASilicon,Zajac2018ResonantlySpins}. In particular Si/SiGe heterostructures are promising for scalable qubit tiles\cite{Vandersypen2017InterfacingCoherent,Lieaar3960} and the presence of low disorder has already made it possible to define a nine quantum dot array\cite{zajac_scalable_2016}. However, spin qubits in silicon suffer from a two-fold degeneracy of the conduction-band valleys\cite{ando_electronic_1982,zwanenburg_silicon_2013,koiller_exchange_2001}, complicating quantum operation. While the valley splitting energy can be large in silicon metal-oxide-semiconductor devices\cite{yang_spin-valley_2013}, even allowing for qubit operation above one Kelvin\cite{yang_operation_2020,petit_universal_2020}, atomic-scale disorder in Si/SiGe heterostructures at the Si quantum well top-interface yields a valley splitting energy that is typically modest and poorly controlled, with values ranging from 10 to 200 $\mu$eV in quantum dots\cite{Watson2018ASilicon,Borselli2011MeasurementDots,hollmann2020large,zajac2015reconfigurable,shi2011tunable,scarlino2017dressed,ferdous2018valley,mi2017high,borjans2019single,mi2018landau}. While Si/SiGe heterostructures may provide a superior host for scalable qubit arrays due to the low disorder, a key challenge is thus to increase the valley splitting energy for scalable quantum information.

The dependence of valley splitting on quantum confinement yields information about the disorder realization at the critical quantum well top-interface and hence provides tools to improve the Si/SiGe platform. The two-dimensional electron gas (2DEG) is confined laterally over the magnetic length scale $l_B=\sqrt{\hbar e/B}$\label{eq:magnlength}, where $B$ is the perpendicular magnetic field, which can be precisely controlled.
The 2DEG is confined vertically by the quantum well heterostructure, with a confinement energy determined by the vertical electric field $E_z$ (perpendicular to the plane of the 2DEG), which pulls the electrons against the top interface. According to the conventional theory, the valley degeneracy is lifted by the broken translational symmetry of the quantum well barriers, and is therefore proportional to the penetration of the wavefunction into the top barrier. This penetration
is proportional to $E_z$ and the two-dimensional electron density~\cite{Friesen2007ValleyWells} $n={\epsilon}E_z/e$, which is easily measured in a Hall bar geometry. However, valley splitting in Si/SiGe 2DEGs is usually probed by activation energy measurements in the quantum Hall regime\cite{weitz1996tilted,lai2004two,Sasaki2009Well-widthWells,Neyens2018TheWells}. In this regime, drawing the correct relationship between valley splitting and electric field is challenging since the presence of quantum Hall edge states adds complexity to the electrostatics of the system compared to the simple electrostatics of an infinite 2DEG. Furthermore, the dependence of valley splitting upon both $B$ and $n$ requires activation energy measurements over many filling factors $\nu$ because of the quantum Hall relationship $\nu=hn/eB$. This has challenged experiments so far, since measurements over many filling factors are possible in heterostructure field effect transistors (H-FETs) only if the mobility is high and the critical density for establishing metallic conduction in the channel (percolation density) is low.

In this Letter we overcome this hurdle and we study valley splitting of 2D electrons as a function of both magnetic field and density in Si/SiGe H-FETs. Benefiting from the high mobility and low percolation density achieved in industrially grown heterostructures\cite{wuetz2019multiplexed}, we resolve Shubnikov--de Haas (SdH) oscillations at small magnetic fields over a large range of densities and we measure activation energies in the quantum Hall regime over an unprecedented range of filling factors. We find that valley splitting increases linearly with magnetic field and is independent of Hall density. Such behavior is inconsistent with bulk transport models; we therefore present a model in which the valley splitting depends on the incremental changes in density $\Delta n = eB/h$ across quantum Hall edge strips. With this critical new insight, the experimental dependence of valley splitting upon $\Delta n$ is in agreement with previous calculations for a near-ideal Si quantum well top-interface\cite{Friesen2007ValleyWells}.

\begin{figure}[htp]
	\includegraphics[width=80mm]{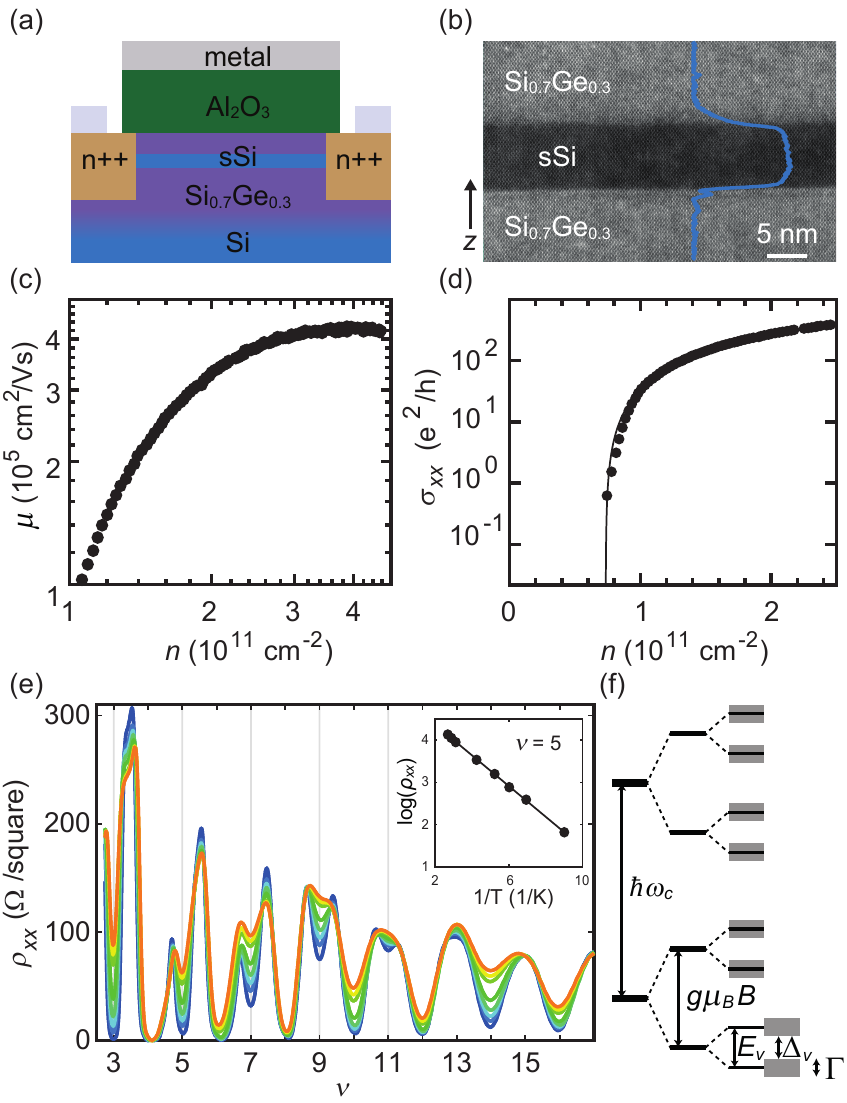}%
	\caption{(a) Cross-section schematic of a Si/SiGe heterostructure field effect transistor. (b) High angle annular dark field scanning transmission electron (HAADF-STEM) image of the strained Si quantum well and nearby Si$_{0.7}$Ge$_{0.3}$ with superimposed HAADF-STEM intensity profile (blue line). The heterostructure growth direction $z$ is indicated by a black arrow (c) Mobility $\mu$ and (d) conductivity $\sigma_{xx}$ as a function of density $n$ at a temperature of 110 mK, measured at the cold finger of the dilution refrigerator. The black line in (d) is a fit to percolation theory. (e) Resistivity $\rho_{xx}$ as a function of filling factor $\nu$ measured at $n$ = 4.0$\times$10$^{11}$cm$^{-2}$.	Different colors correspond to different temperatures from 110 mK (dark blue) to 450 mK (orange). The inset reports the Arrhenius plot and fit to extract $\Delta_v$ for $\nu$ = 5. (f) Single particle Landau level energy diagram. Valley split levels correspond to odd integer filling factors $\nu$, Zeeman split levels to $\nu$ = (4$k$-2) ($k$ = 1,2,3...), whereas spin and valley degenerate Landau levels correspond to $\nu$ = 4$k$. The shaded areas represent the single-particle level broadening $\Gamma$ due to disorder.}
\label{fig:MAT}
\end{figure}

Figure~\ref{fig:MAT} shows the basic structural and magnetotransport characterization of the Si/SiGe H-FETs. The heterostructures were grown by reduced-pressure chemical vapor deposition in an industrial manufacturing CMOS fab on top of a 300 mm Si wafer. The layer sequence [Fig.~\ref{fig:MAT}(a)] comprises a step-graded Si$_{0.7}$Ge$_{0.3}$ strain-relaxed buffer, an 8 nm strained Si quantum well, a 34 nm Si$_{0.7}$Ge$_{0.3}$ barrier, and a sacrificial 3 nm Si cap. Hall-bar shaped H-FETs are fabricated 
with ion implanted ohmic contacts and an Al$_2$O$_3$/Ti/Pt gate stack.
Magnetotransport characterization of
the H-FETs is performed over a temperature range $T$ = 50--500 mK in a dilution refrigerator using standard four-probe low-frequency lock-in techniques. Positive bias applied to the gate induces a 2DEG and controls $n$ in the quantum well
(see Ref.~\cite{wuetz2019multiplexed} for details of the heterostructure growth, device fabrication, and magnetotransport characterization). 
 Figure~\ref{fig:MAT}(b) shows a cross-section image of the heterostructure obtained by high angle annular dark field scanning transmission electron microscopy (HAADF-STEM) to highlight the different chemistry in the layers. By fitting the HAADF-STEM intensity profile in Fig.~\ref{fig:MAT}(b) with an error function\cite{Sammak2019ShallowTechnology}, we infer that the transition between Si and SiGe at the top interface of the quantum well is characterized by a distance $\lambda$ $\approx$ 1 nm\footnote{See Supplemental Material [url] for the analysis of the HAADF-STEM intensity profile along the heterostructure growth direction}. Figure~\ref{fig:MAT}(c) shows the density-dependent mobility. At high density, the mobility is limited by short-range scattering from impurities within or near the quantum well and reaches a maximum value of 4.2$\times10^5$~cm$^2\slash$Vs at $n$ = 4.0$\times$10$^{11}$cm$^{-2}$. A low percolation density of 7.3$\times$10$^{10}$cm$^{-2}$ is extracted by fitting the density-dependent conductivity [Fig.~\ref{fig:MAT}(d)] to percolation theory\cite{Tracy2009ObservationMOSFET}. 
Overall, high mobilities are observed over a wide range of densities, making these H-FETs well suited for quantum Hall measurements over many filling factors. 

 Figure~\ref{fig:MAT}(e) shows typical temperature-dependent measurements of the longitudinal resistivity ($\rho_{xx}$), plotted for clarity against filling factor $\nu$. These measurements are performed at fixed $n$, by keeping the gate voltage constant while sweeping the magnetic field. We observe clear SdH oscillations that are related to the valley splitting $E_v$, the Zeeman splitting $g\mu_BB$, and the cyclotron gap $\hbar\omega_c$ [Fig.~\ref{fig:MAT}(f)]. The inset in Fig.~\ref{fig:MAT}(e) shows a typical temperature dependence of the SdH oscillation minimum for a valley-split level ($\nu$ = 5). We observe a thermally activated dependence $\rho_{xx}\propto\exp{(-\Delta_{v}/2k_BT)}$, from which the mobility gap $\Delta_{v}$ is determined at a specific pair of $B$ and $n$ values satisfying the quantum Hall relationship $\nu=hn/eB$ when $\nu$ is an integer. As indicated in Fig.~\ref{fig:MAT}(f), the mobility gap $\Delta_{v}$ measures the valley splitting $E_{v}$ reduced by $\Gamma$, the Landau level broadening induced by disorder.

\begin{figure}[htp]
	\includegraphics[width=80mm]{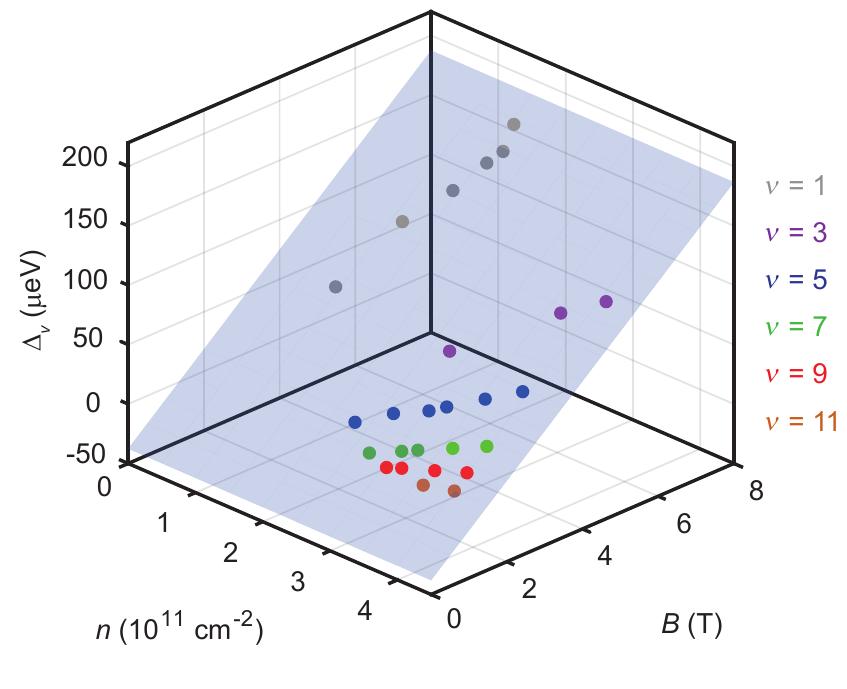}%
	\caption{Activation energy $\Delta_v$ for odd-integer filling factors $\nu$ measured as a function of magnetic field $B$ and Hall density $n$ (filled circles). The blue plane defined by the equation ${\Delta_v=c_BB+c_nn-\Gamma}$ with $c_B=$~28.1~$\mu$eV$/$T, $c_n =$~0.1~$\mu$eV/10$^{11}$cm$^{-2}$, and $\Gamma=$~37.5~$\mu$eV is a fit to the experimental data points with coefficient of determination $R^2=0.993$.}
\label{fig:VS3D}
\end{figure}

Figure~\ref{fig:VS3D} shows $\Delta_v$ as a function of $B$ and $n$ on a three-dimensional (3D) plot. The data points in this graph are obtained by repeating temperature dependent $\rho_{xx}$ measurements at different $n$ and by extracting $\Delta_v$ for the odd-numbered filling factors resolved at each iteration. The 3D plot shows that $\Delta_v$ increases linearly with $B$ and---at fixed $B$---is independent of $n$. These observations are quantified by fitting the data in Fig.~\ref{fig:VS3D} to the plane ${\Delta_v=c_BB+c_nn-\Gamma}$ with coefficient $c_B=28.1 \pm 1.2$ $\mu$eV$/$T, $c_n=0.1 \pm 2.5$ $\mu$eV/10$^{11}$cm$^{-2}$, and $\Gamma=37.5 \pm 10.2$ $\mu$eV. Our main experimental result, $E_v(B,n)=c_BB$, follows by considering $c_n$ negligible and correcting for $\Gamma$\footnote{See Supplemental Material [URL] for theoretical justification of this fitting form}. Under similar experimental conditions we measure a $g$-factor $\approx 1.8$, close to the expected value of 2\footnote{See Supplemental Material [URL] for $g$-factor analysis}. This observation suggests that the measured quantum Hall gaps are not enhanced by electron-electron interactions\cite{Neyens2018TheWells} and that they represent the single particle valley splitting relevant for silicon qubits.

\begin{figure}[htp]
\includegraphics[width=77mm]{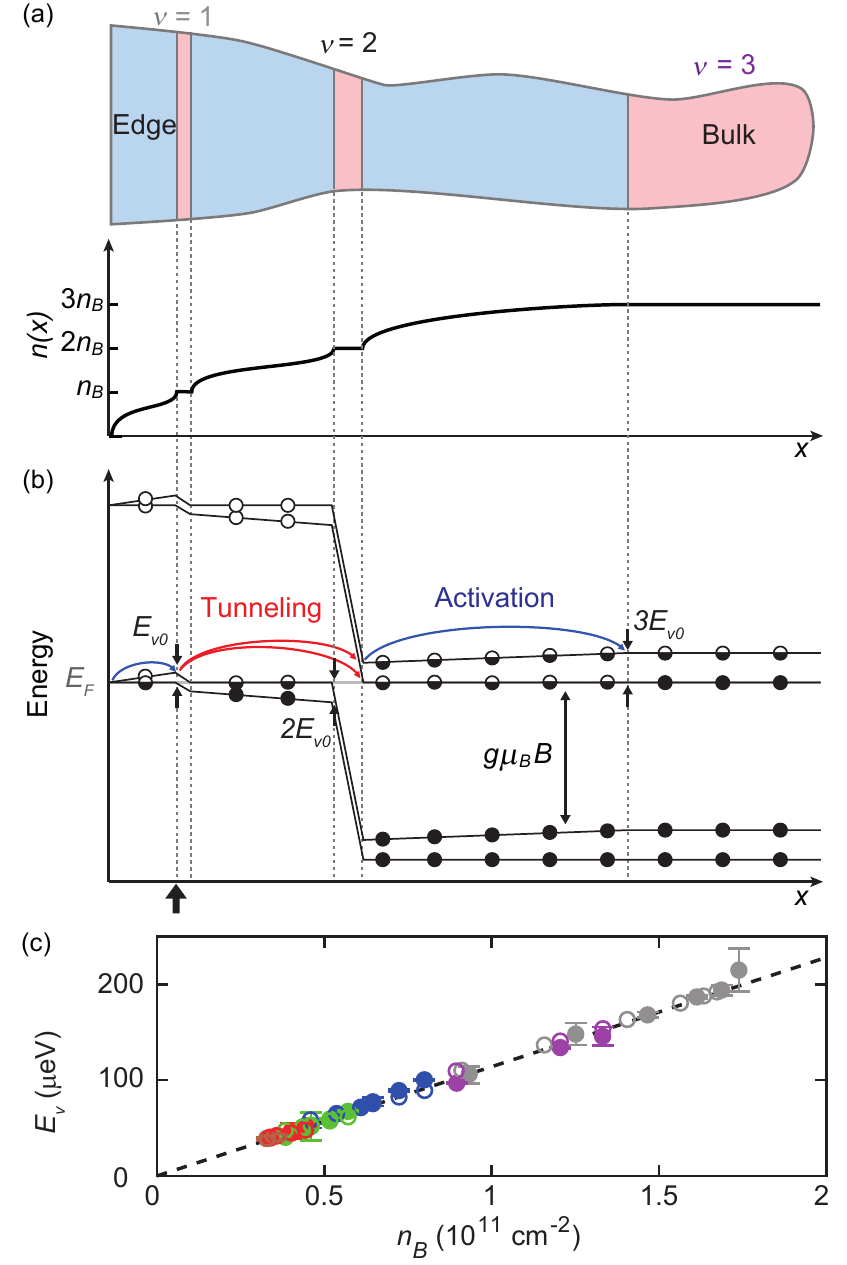}%
\caption{(a) Schematic representation of the charge density profile $n(x)$ on the left-hand side of a Hall bar shaped H-FET for the case of $\nu = 3$, in units of the density $n_B=eB/h$ corresponding to one completely filled Landau level. The edge of the Hall bar is at $x=0$.
The 2DEG is divided into compressible (blue) and incompressible (pink) strips.
(b) Energy-level diagram, including valley and Zeeman splittings. Landau-level splittings are not present for the case of $\nu_\text{bulk}=3$ shown here, but would occur for larger $\nu_\text{bulk}$ values. Valley splittings are assumed to be proportional to the local value of $n$.
Filled, partially filled, and empty Landau levels are indicated by filled, half-filled, and empty circles, respectively. Our model of activated transport incorporates activation and tunneling processes across the alternating compressible and incompressible strips. The thick black arrow indicates the location where the valley splitting takes its characteristic value, $E_{v0}$. The valley splitting increases by an amount $E_{v0}$ in each of the compressible strips.
(c) Agreement between experimental (filled circles) and simulated (open circles) data points of valley splitting $E_v$ as a function of density $n_B = eB/h$. The dashed line is the expected valley splitting dependence on density for a disorder-free quantum well top-interface as calculated in Ref.~\cite{Friesen2007ValleyWells}.}
\label{fig:theory}
\end{figure}

The conventional theory of valley splitting in a silicon quantum well predicts that $E_v$ depends on the penetration of the electron wavefunction into the quantum well barrier, with $E_v\propto E_z$~\cite{Friesen2007ValleyWells}.
If we assume that the 2DEG screens out electric fields from the top gate, then we should find $E_z=0$ at the bottom of the 2DEG and $E_z=en/\epsilon$ at the top, so that $E_v\propto n$, where $n$ is the locally varying electron density in the 2DEG. The proportionality constant is obtained, self-consistently, in Ref.~\cite{Friesen2007ValleyWells}. It is therefore surprising that $E_v$ does not appear to depend on $n$ in the Hall data reported in Fig.~\ref{fig:VS3D}.

Previous experiments on quantum Hall devices were unable to separately determine the dependence of valley splitting on $n$ and $B$.
In particular, there was no indication of behavior inconsistent with conventional ``bulk” behavior.
We must therefore modify previous theories of bulk behavior~\cite{goswami_2007} to account for the fact that valley splitting varies systematically across the device.
Specifically, we propose that the activation energy is determined near the edges of the 2DEG, giving rise to the observed independence of $E_v$ on $n$, as we now explain.

In the quantum Hall regime, the 2DEG forms alternating strips of compressible (blue) and incompressible (pink) liquid~\cite{chklovskii1992electrostatics}, as sketched in Fig.~\ref{fig:theory}(a).
The density increases by $n_B=eB/h$ in consecutive incompressible strips, where $n_B$ is the quantized density of a filled Landau level, until reaching the bulk value $n=\nu_\text{bulk}n_B$, measured by the Hall effect.
In the compressible strips, the density varies monotonically between these quantized values, with a charge distribution that screens out electric fields parallel to the plane of the 2DEG.
In this way, $n$ varies from zero at the edge of the Hall bar to its bulk value in the center.
Figure~\ref{fig:theory}(b) is a sketch of the corresponding energy levels, assuming that $E_v$ is proportional to the local value of $n$.
Note that in the compressible strips and in the bulk, the highest filled levels are pinned at the Fermi level $E_F$~\cite{davies1998physics}.

To observe nonzero longitudinal resistance in our activation energy experiments, electrons must transit across the transverse width of the Hall bar.
However, since all the states in the incompressible strip in the center of the Hall bar are filled for integer filling factors, this requires exciting electrons to a state above the Fermi level.
Our proposed model incorporates alternating activation and tunneling processes across successive compressible strips.
Each of the activation steps involves climbing ``uphill" by an energy $\sim$ $E_{v0}$, which is the change in valley splitting associated with the density change $\Delta n=n_B$.
The tunneling process results in the occupation of two valley states, as indicated, since the valley quantum number is not preserved in the presence of atomic-scale roughness at the quantum-well interface\cite{gamble2013disorder}. This process leads to conduction across the bulk because the valley-state lifetimes are long, so electrons can travel long distances before decaying.
In this model, the characteristic energy $E_{v0}$ is the valley splitting obtained at the position indicated by a thick black arrow in Fig.~\ref{fig:theory}(b).

In Fig.~\ref{fig:theory}(c) we demonstrate the consistency of this model with our experimental results and compare our results with previous effective mass theories for valley splitting in Si/SiGe\cite{Friesen2007ValleyWells}. Here, the experimental results from Fig.~\ref{fig:VS3D} are reported as solid circles as a function of density $n_B = eB/h$. The data points lie on a single line, irrespective of $\nu$, as expected from the discussion of Fig.~\ref{fig:VS3D}. We also report theoretical results for the valley splitting obtained from Thomas-Fermi simulations of the Hall-bar H-FET (open circles\footnote{see Supplemental Material [URL] for theoretical methods, which includes Ref.~\cite{frees2019compressed}}).
In each simulation, we adjust the top-gate voltage to obtain the desired filling factor in the bulk region.
The values of $n$ are chosen to match those used in the experiments (see Fig.~\ref{fig:VS3D}).
Although magnetic field does not enter the simulations explicitly, its value is determined from $n$ and $\nu$ through the quantization relation $B=hn_\text{bulk}/e\nu$.
We then evaluate $E_z$ at the location of the thick black arrow in Fig.~\ref{fig:theory}(c).
Valley splitting is assumed to be proportional to $E_z$ at the top interface of the quantum well, as described above, and we use a single fitting parameter $\beta =$ 134.77 $\mu$eV$\cdot$m/MV to match the simulations with the experimental results, through the relation $E_v = \beta E_z$, correcting for the offset of the experimental data at zero electric field due to $\Gamma$.
The agreement between the experimental and simulated data points indicates that the proposed activation energy model agrees very well with the experimental measurements of quantum Hall gaps. Additionally, we report in Fig.~\ref{fig:theory}(c) as a dashed line the expected value of valley splitting in Si/SiGe according to Eq. 48 of Ref.~\cite{Friesen2007ValleyWells}, which is valid for a near-ideal Si quantum well top-interface. Again, the experimental data matches the theoretical expectations. This result suggests that the atomic-scale disorder associated with the diffused SiGe barrier in Fig.~\ref{fig:MAT}(b) does not significantly suppress valley splitting, at least over lateral length scales less than the largest magnetic confinement length for electrons $\sim 4l_B$ = 70 nm in our experiments.

In summary, we have measured the valley splitting in low-disorder silicon quantum wells over a large range of odd-numbered filling factors in the quantum Hall regime. Supported by a transport model that incorporates the electrostatics of quantum Hall edge states, we demonstrate that valley splitting depends linearly upon the density $eB/h$ rather than on the Hall density. We estimate the ratio $E_v/E_z \sim$ 135 $\mu$eV$\cdot$m/MV, which can be compared directly to valley splitting measurements in quantum dots. 

\vspace{\baselineskip}
This work was supported in part by the Army Research Office (Grant No. W911NF-17-1-0274) and the Vannevar Bush Faculty Fellowship program sponsored by the Basic Research Office of the Assistant Secretary of Defense for Research and Engineering and funded by the Office of Naval Research through Grant No. N00014-15-1-0029. The views and conclusions contained in this document are those of the authors and should not be interpreted as representing the official policies, either expressed or implied, of the Army Research Office (ARO), or the U.S. Government. The U.S. Government is authorized to reproduce and distribute reprints for Government purposes notwithstanding any copyright notation herein. This work is part of the research program OTP with project number 16278, which is (partly) financed by the Netherlands Organisation for Scientific Research (NWO).

\vspace{\baselineskip}
Data sets supporting the findings of this study are available at 10.4121/uuid:46a70228-eb5d-4d41-9d1d-f41 3c1bc0af.

\bibliography{bibliography.bib}

\begin{thebibliography}{41}%
\makeatletter
\providecommand \@ifxundefined [1]{%
 \@ifx{#1\undefined}
}%
\providecommand \@ifnum [1]{%
 \ifnum #1\expandafter \@firstoftwo
 \else \expandafter \@secondoftwo
 \fi
}%
\providecommand \@ifx [1]{%
 \ifx #1\expandafter \@firstoftwo
 \else \expandafter \@secondoftwo
 \fi
}%
\providecommand \natexlab [1]{#1}%
\providecommand \enquote  [1]{``#1''}%
\providecommand \bibnamefont  [1]{#1}%
\providecommand \bibfnamefont [1]{#1}%
\providecommand \citenamefont [1]{#1}%
\providecommand \href@noop [0]{\@secondoftwo}%
\providecommand \href [0]{\begingroup \@sanitize@url \@href}%
\providecommand \@href[1]{\@@startlink{#1}\@@href}%
\providecommand \@@href[1]{\endgroup#1\@@endlink}%
\providecommand \@sanitize@url [0]{\catcode `\\12\catcode `\$12\catcode
  `\&12\catcode `\#12\catcode `\^12\catcode `\_12\catcode `\%12\relax}%
\providecommand \@@startlink[1]{}%
\providecommand \@@endlink[0]{}%
\providecommand \url  [0]{\begingroup\@sanitize@url \@url }%
\providecommand \@url [1]{\endgroup\@href {#1}{\urlprefix }}%
\providecommand \urlprefix  [0]{URL }%
\providecommand \Eprint [0]{\href }%
\providecommand \doibase [0]{http://dx.doi.org/}%
\providecommand \selectlanguage [0]{\@gobble}%
\providecommand \bibinfo  [0]{\@secondoftwo}%
\providecommand \bibfield  [0]{\@secondoftwo}%
\providecommand \translation [1]{[#1]}%
\providecommand \BibitemOpen [0]{}%
\providecommand \bibitemStop [0]{}%
\providecommand \bibitemNoStop [0]{.\EOS\space}%
\providecommand \EOS [0]{\spacefactor3000\relax}%
\providecommand \BibitemShut  [1]{\csname bibitem#1\endcsname}%
\let\auto@bib@innerbib\@empty
%</preamble>
\bibitem [{\citenamefont {Veldhorst}\ \emph {et~al.}(2014)\citenamefont
  {Veldhorst}, \citenamefont {Hwang}, \citenamefont {Yang}, \citenamefont
  {Leenstra}, \citenamefont {de~Ronde}, \citenamefont {Dehollain},
  \citenamefont {Muhonen}, \citenamefont {Hudson}, \citenamefont {Itoh},
  \citenamefont {Morello},\ and\ \citenamefont
  {Dzurak}}]{veldhorst_addressable_2014}%
  \BibitemOpen
  \bibfield  {author} {\bibinfo {author} {\bibfnamefont {M.}~\bibnamefont
  {Veldhorst}}, \bibinfo {author} {\bibfnamefont {J.~C.~C.}\ \bibnamefont
  {Hwang}}, \bibinfo {author} {\bibfnamefont {C.~H.}\ \bibnamefont {Yang}},
  \bibinfo {author} {\bibfnamefont {A.~W.}\ \bibnamefont {Leenstra}}, \bibinfo
  {author} {\bibfnamefont {B.}~\bibnamefont {de~Ronde}}, \bibinfo {author}
  {\bibfnamefont {J.~P.}\ \bibnamefont {Dehollain}}, \bibinfo {author}
  {\bibfnamefont {J.~T.}\ \bibnamefont {Muhonen}}, \bibinfo {author}
  {\bibfnamefont {F.~E.}\ \bibnamefont {Hudson}}, \bibinfo {author}
  {\bibfnamefont {K.~M.}\ \bibnamefont {Itoh}}, \bibinfo {author}
  {\bibfnamefont {A.}~\bibnamefont {Morello}}, \ and\ \bibinfo {author}
  {\bibfnamefont {A.~S.}\ \bibnamefont {Dzurak}},\ }\href {\doibase
  10.1038/nnano.2014.216} {\bibfield  {journal} {\bibinfo  {journal} {Nature
  Nanotechnology}\ }\textbf {\bibinfo {volume} {9}},\ \bibinfo {pages} {981}
  (\bibinfo {year} {2014})}\BibitemShut {NoStop}%
\bibitem [{\citenamefont {Yoneda}\ \emph {et~al.}(2018)\citenamefont {Yoneda},
  \citenamefont {Takeda}, \citenamefont {Otsuka}, \citenamefont {Nakajima},
  \citenamefont {Delbecq}, \citenamefont {Allison}, \citenamefont {Honda},
  \citenamefont {Kodera}, \citenamefont {Oda}, \citenamefont {Hoshi},
  \citenamefont {Usami}, \citenamefont {Itoh},\ and\ \citenamefont
  {Tarucha}}]{Yoneda2018A99.9}%
  \BibitemOpen
  \bibfield  {author} {\bibinfo {author} {\bibfnamefont {J.}~\bibnamefont
  {Yoneda}}, \bibinfo {author} {\bibfnamefont {K.}~\bibnamefont {Takeda}},
  \bibinfo {author} {\bibfnamefont {T.}~\bibnamefont {Otsuka}}, \bibinfo
  {author} {\bibfnamefont {T.}~\bibnamefont {Nakajima}}, \bibinfo {author}
  {\bibfnamefont {M.~R.}\ \bibnamefont {Delbecq}}, \bibinfo {author}
  {\bibfnamefont {G.}~\bibnamefont {Allison}}, \bibinfo {author} {\bibfnamefont
  {T.}~\bibnamefont {Honda}}, \bibinfo {author} {\bibfnamefont
  {T.}~\bibnamefont {Kodera}}, \bibinfo {author} {\bibfnamefont
  {S.}~\bibnamefont {Oda}}, \bibinfo {author} {\bibfnamefont {Y.}~\bibnamefont
  {Hoshi}}, \bibinfo {author} {\bibfnamefont {N.}~\bibnamefont {Usami}},
  \bibinfo {author} {\bibfnamefont {K.~M.}\ \bibnamefont {Itoh}}, \ and\
  \bibinfo {author} {\bibfnamefont {S.}~\bibnamefont {Tarucha}},\ }\href
  {\doibase 10.1038/s41565-017-0014-x} {\bibfield  {journal} {\bibinfo
  {journal} {Nature Nanotechnology}\ }\textbf {\bibinfo {volume} {13}},\
  \bibinfo {pages} {102} (\bibinfo {year} {2018})}\BibitemShut {NoStop}%
\bibitem [{\citenamefont {Yang}\ \emph {et~al.}(2019)\citenamefont {Yang},
  \citenamefont {Chan}, \citenamefont {Harper}, \citenamefont {Huang},
  \citenamefont {Evans}, \citenamefont {Hwang}, \citenamefont {Hensen},
  \citenamefont {Laucht}, \citenamefont {Tanttu}, \citenamefont {Hudson},
  \citenamefont {Flammia}, \citenamefont {Itoh}, \citenamefont {Morello},
  \citenamefont {Bartlett},\ and\ \citenamefont {Dzurak}}]{yang_silicon_2019}%
  \BibitemOpen
  \bibfield  {author} {\bibinfo {author} {\bibfnamefont {C.~H.}\ \bibnamefont
  {Yang}}, \bibinfo {author} {\bibfnamefont {K.~W.}\ \bibnamefont {Chan}},
  \bibinfo {author} {\bibfnamefont {R.}~\bibnamefont {Harper}}, \bibinfo
  {author} {\bibfnamefont {W.}~\bibnamefont {Huang}}, \bibinfo {author}
  {\bibfnamefont {T.}~\bibnamefont {Evans}}, \bibinfo {author} {\bibfnamefont
  {J.~C.~C.}\ \bibnamefont {Hwang}}, \bibinfo {author} {\bibfnamefont
  {B.}~\bibnamefont {Hensen}}, \bibinfo {author} {\bibfnamefont
  {A.}~\bibnamefont {Laucht}}, \bibinfo {author} {\bibfnamefont
  {T.}~\bibnamefont {Tanttu}}, \bibinfo {author} {\bibfnamefont {F.~E.}\
  \bibnamefont {Hudson}}, \bibinfo {author} {\bibfnamefont {S.~T.}\
  \bibnamefont {Flammia}}, \bibinfo {author} {\bibfnamefont {K.~M.}\
  \bibnamefont {Itoh}}, \bibinfo {author} {\bibfnamefont {A.}~\bibnamefont
  {Morello}}, \bibinfo {author} {\bibfnamefont {S.~D.}\ \bibnamefont
  {Bartlett}}, \ and\ \bibinfo {author} {\bibfnamefont {A.~S.}\ \bibnamefont
  {Dzurak}},\ }\href {\doibase 10.1038/s41928-019-0234-1} {\bibfield  {journal}
  {\bibinfo  {journal} {Nature Electronics}\ }\textbf {\bibinfo {volume} {2}},\
  \bibinfo {pages} {151} (\bibinfo {year} {2019})}\BibitemShut {NoStop}%
\bibitem [{\citenamefont {Veldhorst}\ \emph {et~al.}(2015)\citenamefont
  {Veldhorst}, \citenamefont {Yang}, \citenamefont {Hwang}, \citenamefont
  {Huang}, \citenamefont {Dehollain}, \citenamefont {Muhonen}, \citenamefont
  {Simmons}, \citenamefont {Laucht}, \citenamefont {Hudson}, \citenamefont
  {Itoh}, \citenamefont {Morello},\ and\ \citenamefont
  {Dzurak}}]{veldhorst_two-qubit_2015}%
  \BibitemOpen
  \bibfield  {author} {\bibinfo {author} {\bibfnamefont {M.}~\bibnamefont
  {Veldhorst}}, \bibinfo {author} {\bibfnamefont {C.~H.}\ \bibnamefont {Yang}},
  \bibinfo {author} {\bibfnamefont {J.~C.~C.}\ \bibnamefont {Hwang}}, \bibinfo
  {author} {\bibfnamefont {W.}~\bibnamefont {Huang}}, \bibinfo {author}
  {\bibfnamefont {J.~P.}\ \bibnamefont {Dehollain}}, \bibinfo {author}
  {\bibfnamefont {J.~T.}\ \bibnamefont {Muhonen}}, \bibinfo {author}
  {\bibfnamefont {S.}~\bibnamefont {Simmons}}, \bibinfo {author} {\bibfnamefont
  {A.}~\bibnamefont {Laucht}}, \bibinfo {author} {\bibfnamefont {F.~E.}\
  \bibnamefont {Hudson}}, \bibinfo {author} {\bibfnamefont {K.~M.}\
  \bibnamefont {Itoh}}, \bibinfo {author} {\bibfnamefont {A.}~\bibnamefont
  {Morello}}, \ and\ \bibinfo {author} {\bibfnamefont {A.~S.}\ \bibnamefont
  {Dzurak}},\ }\href {\doibase 10.1038/nature15263} {\bibfield  {journal}
  {\bibinfo  {journal} {Nature}\ }\textbf {\bibinfo {volume} {526}},\ \bibinfo
  {pages} {410} (\bibinfo {year} {2015})}\BibitemShut {NoStop}%
\bibitem [{\citenamefont {Watson}\ \emph {et~al.}(2018)\citenamefont {Watson},
  \citenamefont {Philips}, \citenamefont {Kawakami}, \citenamefont {Ward},
  \citenamefont {Scarlino}, \citenamefont {Veldhorst}, \citenamefont {Savage},
  \citenamefont {Lagally}, \citenamefont {Friesen}, \citenamefont
  {Coppersmith}, \citenamefont {Eriksson},\ and\ \citenamefont
  {Vandersypen}}]{Watson2018ASilicon}%
  \BibitemOpen
  \bibfield  {author} {\bibinfo {author} {\bibfnamefont {T.~F.}\ \bibnamefont
  {Watson}}, \bibinfo {author} {\bibfnamefont {S.~G.}\ \bibnamefont {Philips}},
  \bibinfo {author} {\bibfnamefont {E.}~\bibnamefont {Kawakami}}, \bibinfo
  {author} {\bibfnamefont {D.~R.}\ \bibnamefont {Ward}}, \bibinfo {author}
  {\bibfnamefont {P.}~\bibnamefont {Scarlino}}, \bibinfo {author}
  {\bibfnamefont {M.}~\bibnamefont {Veldhorst}}, \bibinfo {author}
  {\bibfnamefont {D.~E.}\ \bibnamefont {Savage}}, \bibinfo {author}
  {\bibfnamefont {M.~G.}\ \bibnamefont {Lagally}}, \bibinfo {author}
  {\bibfnamefont {M.}~\bibnamefont {Friesen}}, \bibinfo {author} {\bibfnamefont
  {S.~N.}\ \bibnamefont {Coppersmith}}, \bibinfo {author} {\bibfnamefont
  {M.~A.}\ \bibnamefont {Eriksson}}, \ and\ \bibinfo {author} {\bibfnamefont
  {L.~M.}\ \bibnamefont {Vandersypen}},\ }\href {\doibase 10.1038/nature25766}
  {\bibfield  {journal} {\bibinfo  {journal} {Nature}\ }\textbf {\bibinfo
  {volume} {555}},\ \bibinfo {pages} {633} (\bibinfo {year}
  {2018})}\BibitemShut {NoStop}%
\bibitem [{\citenamefont {Zajac}\ \emph {et~al.}(2018)\citenamefont {Zajac},
  \citenamefont {Sigillito}, \citenamefont {Russ}, \citenamefont {Borjans},
  \citenamefont {Taylor}, \citenamefont {Burkard},\ and\ \citenamefont
  {Petta}}]{Zajac2018ResonantlySpins}%
  \BibitemOpen
  \bibfield  {author} {\bibinfo {author} {\bibfnamefont {D.~M.}\ \bibnamefont
  {Zajac}}, \bibinfo {author} {\bibfnamefont {A.~J.}\ \bibnamefont
  {Sigillito}}, \bibinfo {author} {\bibfnamefont {M.}~\bibnamefont {Russ}},
  \bibinfo {author} {\bibfnamefont {F.}~\bibnamefont {Borjans}}, \bibinfo
  {author} {\bibfnamefont {J.~M.}\ \bibnamefont {Taylor}}, \bibinfo {author}
  {\bibfnamefont {G.}~\bibnamefont {Burkard}}, \ and\ \bibinfo {author}
  {\bibfnamefont {J.~R.}\ \bibnamefont {Petta}},\ }\href {\doibase
  10.1126/science.aao5965} {\bibfield  {journal} {\bibinfo  {journal}
  {Science}\ }\textbf {\bibinfo {volume} {359}},\ \bibinfo {pages} {439}
  (\bibinfo {year} {2018})}\BibitemShut {NoStop}%
\bibitem [{\citenamefont {Vandersypen}\ \emph {et~al.}(2017)\citenamefont
  {Vandersypen}, \citenamefont {Bluhm}, \citenamefont {Clarke}, \citenamefont
  {Dzurak}, \citenamefont {Ishihara}, \citenamefont {Morello}, \citenamefont
  {Reilly}, \citenamefont {Schreiber},\ and\ \citenamefont
  {Veldhorst}}]{Vandersypen2017InterfacingCoherent}%
  \BibitemOpen
  \bibfield  {author} {\bibinfo {author} {\bibfnamefont {L.~M.~K.}\
  \bibnamefont {Vandersypen}}, \bibinfo {author} {\bibfnamefont
  {H.}~\bibnamefont {Bluhm}}, \bibinfo {author} {\bibfnamefont {J.~S.}\
  \bibnamefont {Clarke}}, \bibinfo {author} {\bibfnamefont {A.~S.}\
  \bibnamefont {Dzurak}}, \bibinfo {author} {\bibfnamefont {R.}~\bibnamefont
  {Ishihara}}, \bibinfo {author} {\bibfnamefont {A.}~\bibnamefont {Morello}},
  \bibinfo {author} {\bibfnamefont {D.~J.}\ \bibnamefont {Reilly}}, \bibinfo
  {author} {\bibfnamefont {L.~R.}\ \bibnamefont {Schreiber}}, \ and\ \bibinfo
  {author} {\bibfnamefont {M.}~\bibnamefont {Veldhorst}},\ }\href {\doibase
  10.1038/s41534-017-0038-y} {\bibfield  {journal} {\bibinfo  {journal} {npj
  Quantum Information}\ }\textbf {\bibinfo {volume} {3}},\ \bibinfo {pages}
  {34} (\bibinfo {year} {2017})}\BibitemShut {NoStop}%
\bibitem [{\citenamefont {Li}\ \emph {et~al.}(2018)\citenamefont {Li},
  \citenamefont {Petit}, \citenamefont {Franke}, \citenamefont {Dehollain},
  \citenamefont {Helsen}, \citenamefont {Steudtner}, \citenamefont {Thomas},
  \citenamefont {Yoscovits}, \citenamefont {Singh}, \citenamefont {Wehner},
  \citenamefont {Vandersypen}, \citenamefont {Clarke},\ and\ \citenamefont
  {Veldhorst}}]{Lieaar3960}%
  \BibitemOpen
  \bibfield  {author} {\bibinfo {author} {\bibfnamefont {R.}~\bibnamefont
  {Li}}, \bibinfo {author} {\bibfnamefont {L.}~\bibnamefont {Petit}}, \bibinfo
  {author} {\bibfnamefont {D.~P.}\ \bibnamefont {Franke}}, \bibinfo {author}
  {\bibfnamefont {J.~P.}\ \bibnamefont {Dehollain}}, \bibinfo {author}
  {\bibfnamefont {J.}~\bibnamefont {Helsen}}, \bibinfo {author} {\bibfnamefont
  {M.}~\bibnamefont {Steudtner}}, \bibinfo {author} {\bibfnamefont {N.~K.}\
  \bibnamefont {Thomas}}, \bibinfo {author} {\bibfnamefont {Z.~R.}\
  \bibnamefont {Yoscovits}}, \bibinfo {author} {\bibfnamefont {K.~J.}\
  \bibnamefont {Singh}}, \bibinfo {author} {\bibfnamefont {S.}~\bibnamefont
  {Wehner}}, \bibinfo {author} {\bibfnamefont {L.~M.~K.}\ \bibnamefont
  {Vandersypen}}, \bibinfo {author} {\bibfnamefont {J.~S.}\ \bibnamefont
  {Clarke}}, \ and\ \bibinfo {author} {\bibfnamefont {M.}~\bibnamefont
  {Veldhorst}},\ }\href@noop {} {\bibfield  {journal} {\bibinfo  {journal}
  {Science Advances}\ }\textbf {\bibinfo {volume} {4}},\ \bibinfo {pages}
  {eaar3960} (\bibinfo {year} {2018})}\BibitemShut {NoStop}%
\bibitem [{\citenamefont {Zajac}\ \emph {et~al.}(2016)\citenamefont {Zajac},
  \citenamefont {Hazard}, \citenamefont {Mi}, \citenamefont {Nielsen},\ and\
  \citenamefont {Petta}}]{zajac_scalable_2016}%
  \BibitemOpen
  \bibfield  {author} {\bibinfo {author} {\bibfnamefont {D.}~\bibnamefont
  {Zajac}}, \bibinfo {author} {\bibfnamefont {T.}~\bibnamefont {Hazard}},
  \bibinfo {author} {\bibfnamefont {X.}~\bibnamefont {Mi}}, \bibinfo {author}
  {\bibfnamefont {E.}~\bibnamefont {Nielsen}}, \ and\ \bibinfo {author}
  {\bibfnamefont {J.}~\bibnamefont {Petta}},\ }\href {\doibase
  10.1103/PhysRevApplied.6.054013} {\bibfield  {journal} {\bibinfo  {journal}
  {Physical Review Applied}\ }\textbf {\bibinfo {volume} {6}},\ \bibinfo
  {pages} {054013} (\bibinfo {year} {2016})}\BibitemShut {NoStop}%
\bibitem [{\citenamefont {Ando}\ \emph {et~al.}(1982)\citenamefont {Ando},
  \citenamefont {Fowler},\ and\ \citenamefont {Stern}}]{ando_electronic_1982}%
  \BibitemOpen
  \bibfield  {author} {\bibinfo {author} {\bibfnamefont {T.}~\bibnamefont
  {Ando}}, \bibinfo {author} {\bibfnamefont {A.~B.}\ \bibnamefont {Fowler}}, \
  and\ \bibinfo {author} {\bibfnamefont {F.}~\bibnamefont {Stern}},\ }\href
  {\doibase 10.1103/RevModPhys.54.437} {\bibfield  {journal} {\bibinfo
  {journal} {Reviews of Modern Physics}\ }\textbf {\bibinfo {volume} {54}},\
  \bibinfo {pages} {437} (\bibinfo {year} {1982})}\BibitemShut {NoStop}%
\bibitem [{\citenamefont {Zwanenburg}\ \emph {et~al.}(2013)\citenamefont
  {Zwanenburg}, \citenamefont {Dzurak}, \citenamefont {Morello}, \citenamefont
  {Simmons}, \citenamefont {Hollenberg}, \citenamefont {Klimeck}, \citenamefont
  {Rogge}, \citenamefont {Coppersmith},\ and\ \citenamefont
  {Eriksson}}]{zwanenburg_silicon_2013}%
  \BibitemOpen
  \bibfield  {author} {\bibinfo {author} {\bibfnamefont {F.~A.}\ \bibnamefont
  {Zwanenburg}}, \bibinfo {author} {\bibfnamefont {A.~S.}\ \bibnamefont
  {Dzurak}}, \bibinfo {author} {\bibfnamefont {A.}~\bibnamefont {Morello}},
  \bibinfo {author} {\bibfnamefont {M.~Y.}\ \bibnamefont {Simmons}}, \bibinfo
  {author} {\bibfnamefont {L.~C.~L.}\ \bibnamefont {Hollenberg}}, \bibinfo
  {author} {\bibfnamefont {G.}~\bibnamefont {Klimeck}}, \bibinfo {author}
  {\bibfnamefont {S.}~\bibnamefont {Rogge}}, \bibinfo {author} {\bibfnamefont
  {S.~N.}\ \bibnamefont {Coppersmith}}, \ and\ \bibinfo {author} {\bibfnamefont
  {M.~A.}\ \bibnamefont {Eriksson}},\ }\href {\doibase
  10.1103/RevModPhys.85.961} {\bibfield  {journal} {\bibinfo  {journal}
  {Reviews of Modern Physics}\ }\textbf {\bibinfo {volume} {85}},\ \bibinfo
  {pages} {961} (\bibinfo {year} {2013})}\BibitemShut {NoStop}%
\bibitem [{\citenamefont {Koiller}\ \emph {et~al.}(2001)\citenamefont
  {Koiller}, \citenamefont {Hu},\ and\ \citenamefont
  {Das~Sarma}}]{koiller_exchange_2001}%
  \BibitemOpen
  \bibfield  {author} {\bibinfo {author} {\bibfnamefont {B.}~\bibnamefont
  {Koiller}}, \bibinfo {author} {\bibfnamefont {X.}~\bibnamefont {Hu}}, \ and\
  \bibinfo {author} {\bibfnamefont {S.}~\bibnamefont {Das~Sarma}},\ }\href
  {\doibase 10.1103/PhysRevLett.88.027903} {\bibfield  {journal} {\bibinfo
  {journal} {Physical Review Letters}\ }\textbf {\bibinfo {volume} {88}},\
  \bibinfo {pages} {027903} (\bibinfo {year} {2001})}\BibitemShut {NoStop}%
\bibitem [{\citenamefont {Yang}\ \emph {et~al.}(2013)\citenamefont {Yang},
  \citenamefont {Rossi}, \citenamefont {Ruskov}, \citenamefont {Lai},
  \citenamefont {Mohiyaddin}, \citenamefont {Lee}, \citenamefont {Tahan},
  \citenamefont {Klimeck}, \citenamefont {Morello},\ and\ \citenamefont
  {Dzurak}}]{yang_spin-valley_2013}%
  \BibitemOpen
  \bibfield  {author} {\bibinfo {author} {\bibfnamefont {C.~H.}\ \bibnamefont
  {Yang}}, \bibinfo {author} {\bibfnamefont {A.}~\bibnamefont {Rossi}},
  \bibinfo {author} {\bibfnamefont {R.}~\bibnamefont {Ruskov}}, \bibinfo
  {author} {\bibfnamefont {N.~S.}\ \bibnamefont {Lai}}, \bibinfo {author}
  {\bibfnamefont {F.~A.}\ \bibnamefont {Mohiyaddin}}, \bibinfo {author}
  {\bibfnamefont {S.}~\bibnamefont {Lee}}, \bibinfo {author} {\bibfnamefont
  {C.}~\bibnamefont {Tahan}}, \bibinfo {author} {\bibfnamefont
  {G.}~\bibnamefont {Klimeck}}, \bibinfo {author} {\bibfnamefont
  {A.}~\bibnamefont {Morello}}, \ and\ \bibinfo {author} {\bibfnamefont
  {A.~S.}\ \bibnamefont {Dzurak}},\ }\href {\doibase 10.1038/ncomms3069}
  {\bibfield  {journal} {\bibinfo  {journal} {Nature Communications}\ }\textbf
  {\bibinfo {volume} {4}},\ \bibinfo {pages} {2069} (\bibinfo {year}
  {2013})}\BibitemShut {NoStop}%
\bibitem [{\citenamefont {Yang}\ \emph {et~al.}(2020)\citenamefont {Yang},
  \citenamefont {Leon}, \citenamefont {Hwang}, \citenamefont {Saraiva},
  \citenamefont {Tanttu}, \citenamefont {Huang}, \citenamefont
  {Camirand~Lemyre}, \citenamefont {Chan}, \citenamefont {Tan}, \citenamefont
  {Hudson}, \citenamefont {Itoh}, \citenamefont {Morello}, \citenamefont
  {Pioro-Ladrière}, \citenamefont {Laucht},\ and\ \citenamefont
  {Dzurak}}]{yang_operation_2020}%
  \BibitemOpen
  \bibfield  {author} {\bibinfo {author} {\bibfnamefont {C.~H.}\ \bibnamefont
  {Yang}}, \bibinfo {author} {\bibfnamefont {R.~C.~C.}\ \bibnamefont {Leon}},
  \bibinfo {author} {\bibfnamefont {J.~C.~C.}\ \bibnamefont {Hwang}}, \bibinfo
  {author} {\bibfnamefont {A.}~\bibnamefont {Saraiva}}, \bibinfo {author}
  {\bibfnamefont {T.}~\bibnamefont {Tanttu}}, \bibinfo {author} {\bibfnamefont
  {W.}~\bibnamefont {Huang}}, \bibinfo {author} {\bibfnamefont
  {J.}~\bibnamefont {Camirand~Lemyre}}, \bibinfo {author} {\bibfnamefont
  {K.~W.}\ \bibnamefont {Chan}}, \bibinfo {author} {\bibfnamefont {K.~Y.}\
  \bibnamefont {Tan}}, \bibinfo {author} {\bibfnamefont {F.~E.}\ \bibnamefont
  {Hudson}}, \bibinfo {author} {\bibfnamefont {K.~M.}\ \bibnamefont {Itoh}},
  \bibinfo {author} {\bibfnamefont {A.}~\bibnamefont {Morello}}, \bibinfo
  {author} {\bibfnamefont {M.}~\bibnamefont {Pioro-Ladrière}}, \bibinfo
  {author} {\bibfnamefont {A.}~\bibnamefont {Laucht}}, \ and\ \bibinfo {author}
  {\bibfnamefont {A.~S.}\ \bibnamefont {Dzurak}},\ }\href {\doibase
  10.1038/s41586-020-2171-6} {\bibfield  {journal} {\bibinfo  {journal}
  {Nature}\ }\textbf {\bibinfo {volume} {580}},\ \bibinfo {pages} {350}
  (\bibinfo {year} {2020})}\BibitemShut {NoStop}%
\bibitem [{\citenamefont {Petit}\ \emph {et~al.}(2020)\citenamefont {Petit},
  \citenamefont {Eenink}, \citenamefont {Russ}, \citenamefont {Lawrie},
  \citenamefont {Hendrickx}, \citenamefont {Philips}, \citenamefont {Clarke},
  \citenamefont {Vandersypen},\ and\ \citenamefont
  {Veldhorst}}]{petit_universal_2020}%
  \BibitemOpen
  \bibfield  {author} {\bibinfo {author} {\bibfnamefont {L.}~\bibnamefont
  {Petit}}, \bibinfo {author} {\bibfnamefont {H.~G.~J.}\ \bibnamefont
  {Eenink}}, \bibinfo {author} {\bibfnamefont {M.}~\bibnamefont {Russ}},
  \bibinfo {author} {\bibfnamefont {W.~I.~L.}\ \bibnamefont {Lawrie}}, \bibinfo
  {author} {\bibfnamefont {N.~W.}\ \bibnamefont {Hendrickx}}, \bibinfo {author}
  {\bibfnamefont {S.~G.~J.}\ \bibnamefont {Philips}}, \bibinfo {author}
  {\bibfnamefont {J.~S.}\ \bibnamefont {Clarke}}, \bibinfo {author}
  {\bibfnamefont {L.~M.~K.}\ \bibnamefont {Vandersypen}}, \ and\ \bibinfo
  {author} {\bibfnamefont {M.}~\bibnamefont {Veldhorst}},\ }\href {\doibase
  10.1038/s41586-020-2170-7} {\bibfield  {journal} {\bibinfo  {journal}
  {Nature}\ }\textbf {\bibinfo {volume} {580}},\ \bibinfo {pages} {355}
  (\bibinfo {year} {2020})}\BibitemShut {NoStop}%
\bibitem [{\citenamefont {Borselli}\ \emph {et~al.}(2011)\citenamefont
  {Borselli}, \citenamefont {Ross}, \citenamefont {Kiselev}, \citenamefont
  {Croke}, \citenamefont {Holabird}, \citenamefont {Deelman}, \citenamefont
  {Warren}, \citenamefont {Alvarado-Rodriguez}, \citenamefont {Milosavljevic},
  \citenamefont {Ku}, \citenamefont {Wong}, \citenamefont {Schmitz},
  \citenamefont {Sokolich}, \citenamefont {Gyure},\ and\ \citenamefont
  {Hunter}}]{Borselli2011MeasurementDots}%
  \BibitemOpen
  \bibfield  {author} {\bibinfo {author} {\bibfnamefont {M.~G.}\ \bibnamefont
  {Borselli}}, \bibinfo {author} {\bibfnamefont {R.~S.}\ \bibnamefont {Ross}},
  \bibinfo {author} {\bibfnamefont {A.~A.}\ \bibnamefont {Kiselev}}, \bibinfo
  {author} {\bibfnamefont {E.~T.}\ \bibnamefont {Croke}}, \bibinfo {author}
  {\bibfnamefont {K.~S.}\ \bibnamefont {Holabird}}, \bibinfo {author}
  {\bibfnamefont {P.~W.}\ \bibnamefont {Deelman}}, \bibinfo {author}
  {\bibfnamefont {L.~D.}\ \bibnamefont {Warren}}, \bibinfo {author}
  {\bibfnamefont {I.}~\bibnamefont {Alvarado-Rodriguez}}, \bibinfo {author}
  {\bibfnamefont {I.}~\bibnamefont {Milosavljevic}}, \bibinfo {author}
  {\bibfnamefont {F.~C.}\ \bibnamefont {Ku}}, \bibinfo {author} {\bibfnamefont
  {W.~S.}\ \bibnamefont {Wong}}, \bibinfo {author} {\bibfnamefont {A.~E.}\
  \bibnamefont {Schmitz}}, \bibinfo {author} {\bibfnamefont {M.}~\bibnamefont
  {Sokolich}}, \bibinfo {author} {\bibfnamefont {M.~F.}\ \bibnamefont {Gyure}},
  \ and\ \bibinfo {author} {\bibfnamefont {A.~T.}\ \bibnamefont {Hunter}},\
  }\href@noop {} {\bibfield  {journal} {\bibinfo  {journal} {Applied Physics
  Letters}\ }\textbf {\bibinfo {volume} {98}},\ \bibinfo {pages} {123118}
  (\bibinfo {year} {2011})}\BibitemShut {NoStop}%
\bibitem [{\citenamefont {Hollmann}\ \emph {et~al.}(2020)\citenamefont
  {Hollmann}, \citenamefont {Struck}, \citenamefont {Langrock}, \citenamefont
  {Schmidbauer}, \citenamefont {Schauer}, \citenamefont {Leonhardt},
  \citenamefont {Sawano}, \citenamefont {Riemann}, \citenamefont {Abrosimov},
  \citenamefont {Bougeard} \emph {et~al.}}]{hollmann2020large}%
  \BibitemOpen
  \bibfield  {author} {\bibinfo {author} {\bibfnamefont {A.}~\bibnamefont
  {Hollmann}}, \bibinfo {author} {\bibfnamefont {T.}~\bibnamefont {Struck}},
  \bibinfo {author} {\bibfnamefont {V.}~\bibnamefont {Langrock}}, \bibinfo
  {author} {\bibfnamefont {A.}~\bibnamefont {Schmidbauer}}, \bibinfo {author}
  {\bibfnamefont {F.}~\bibnamefont {Schauer}}, \bibinfo {author} {\bibfnamefont
  {T.}~\bibnamefont {Leonhardt}}, \bibinfo {author} {\bibfnamefont
  {K.}~\bibnamefont {Sawano}}, \bibinfo {author} {\bibfnamefont
  {H.}~\bibnamefont {Riemann}}, \bibinfo {author} {\bibfnamefont {N.~V.}\
  \bibnamefont {Abrosimov}}, \bibinfo {author} {\bibfnamefont {D.}~\bibnamefont
  {Bougeard}},  \emph {et~al.},\ }\href@noop {} {\bibfield  {journal} {\bibinfo
   {journal} {Physical Review Applied}\ }\textbf {\bibinfo {volume} {13}},\
  \bibinfo {pages} {034068} (\bibinfo {year} {2020})}\BibitemShut {NoStop}%
\bibitem [{\citenamefont {Zajac}\ \emph {et~al.}(2015)\citenamefont {Zajac},
  \citenamefont {Hazard}, \citenamefont {Mi}, \citenamefont {Wang},\ and\
  \citenamefont {Petta}}]{zajac2015reconfigurable}%
  \BibitemOpen
  \bibfield  {author} {\bibinfo {author} {\bibfnamefont {D.}~\bibnamefont
  {Zajac}}, \bibinfo {author} {\bibfnamefont {T.}~\bibnamefont {Hazard}},
  \bibinfo {author} {\bibfnamefont {X.}~\bibnamefont {Mi}}, \bibinfo {author}
  {\bibfnamefont {K.}~\bibnamefont {Wang}}, \ and\ \bibinfo {author}
  {\bibfnamefont {J.~R.}\ \bibnamefont {Petta}},\ }\href@noop {} {\bibfield
  {journal} {\bibinfo  {journal} {Applied Physics Letters}\ }\textbf {\bibinfo
  {volume} {106}},\ \bibinfo {pages} {223507} (\bibinfo {year}
  {2015})}\BibitemShut {NoStop}%
\bibitem [{\citenamefont {Shi}\ \emph {et~al.}(2011)\citenamefont {Shi},
  \citenamefont {Simmons}, \citenamefont {Prance}, \citenamefont {King~Gamble},
  \citenamefont {Friesen}, \citenamefont {Savage}, \citenamefont {Lagally},
  \citenamefont {Coppersmith},\ and\ \citenamefont
  {Eriksson}}]{shi2011tunable}%
  \BibitemOpen
  \bibfield  {author} {\bibinfo {author} {\bibfnamefont {Z.}~\bibnamefont
  {Shi}}, \bibinfo {author} {\bibfnamefont {C.}~\bibnamefont {Simmons}},
  \bibinfo {author} {\bibfnamefont {J.}~\bibnamefont {Prance}}, \bibinfo
  {author} {\bibfnamefont {J.}~\bibnamefont {King~Gamble}}, \bibinfo {author}
  {\bibfnamefont {M.}~\bibnamefont {Friesen}}, \bibinfo {author} {\bibfnamefont
  {D.}~\bibnamefont {Savage}}, \bibinfo {author} {\bibfnamefont
  {M.}~\bibnamefont {Lagally}}, \bibinfo {author} {\bibfnamefont
  {S.}~\bibnamefont {Coppersmith}}, \ and\ \bibinfo {author} {\bibfnamefont
  {M.}~\bibnamefont {Eriksson}},\ }\href@noop {} {\bibfield  {journal}
  {\bibinfo  {journal} {Applied Physics Letters}\ }\textbf {\bibinfo {volume}
  {99}},\ \bibinfo {pages} {233108} (\bibinfo {year} {2011})}\BibitemShut
  {NoStop}%
\bibitem [{\citenamefont {Scarlino}\ \emph {et~al.}(2017)\citenamefont
  {Scarlino}, \citenamefont {Kawakami}, \citenamefont {Jullien}, \citenamefont
  {Ward}, \citenamefont {Savage}, \citenamefont {Lagally}, \citenamefont
  {Friesen}, \citenamefont {Coppersmith}, \citenamefont {Eriksson},\ and\
  \citenamefont {Vandersypen}}]{scarlino2017dressed}%
  \BibitemOpen
  \bibfield  {author} {\bibinfo {author} {\bibfnamefont {P.}~\bibnamefont
  {Scarlino}}, \bibinfo {author} {\bibfnamefont {E.}~\bibnamefont {Kawakami}},
  \bibinfo {author} {\bibfnamefont {T.}~\bibnamefont {Jullien}}, \bibinfo
  {author} {\bibfnamefont {D.~R.}\ \bibnamefont {Ward}}, \bibinfo {author}
  {\bibfnamefont {D.~E.}\ \bibnamefont {Savage}}, \bibinfo {author}
  {\bibfnamefont {M.~G.}\ \bibnamefont {Lagally}}, \bibinfo {author}
  {\bibfnamefont {M.}~\bibnamefont {Friesen}}, \bibinfo {author} {\bibfnamefont
  {S.~N.}\ \bibnamefont {Coppersmith}}, \bibinfo {author} {\bibfnamefont
  {M.~A.}\ \bibnamefont {Eriksson}}, \ and\ \bibinfo {author} {\bibfnamefont
  {L.~M.~K.}\ \bibnamefont {Vandersypen}},\ }\href {\doibase
  10.1103/PhysRevB.95.165429} {\bibfield  {journal} {\bibinfo  {journal} {Phys.
  Rev. B}\ }\textbf {\bibinfo {volume} {95}},\ \bibinfo {pages} {165429}
  (\bibinfo {year} {2017})}\BibitemShut {NoStop}%
\bibitem [{\citenamefont {Ferdous}\ \emph {et~al.}(2018)\citenamefont
  {Ferdous}, \citenamefont {Kawakami}, \citenamefont {Scarlino}, \citenamefont
  {Nowak}, \citenamefont {Ward}, \citenamefont {Savage}, \citenamefont
  {Lagally}, \citenamefont {Coppersmith}, \citenamefont {Friesen},
  \citenamefont {Eriksson} \emph {et~al.}}]{ferdous2018valley}%
  \BibitemOpen
  \bibfield  {author} {\bibinfo {author} {\bibfnamefont {R.}~\bibnamefont
  {Ferdous}}, \bibinfo {author} {\bibfnamefont {E.}~\bibnamefont {Kawakami}},
  \bibinfo {author} {\bibfnamefont {P.}~\bibnamefont {Scarlino}}, \bibinfo
  {author} {\bibfnamefont {M.~P.}\ \bibnamefont {Nowak}}, \bibinfo {author}
  {\bibfnamefont {D.}~\bibnamefont {Ward}}, \bibinfo {author} {\bibfnamefont
  {D.}~\bibnamefont {Savage}}, \bibinfo {author} {\bibfnamefont
  {M.}~\bibnamefont {Lagally}}, \bibinfo {author} {\bibfnamefont
  {S.}~\bibnamefont {Coppersmith}}, \bibinfo {author} {\bibfnamefont
  {M.}~\bibnamefont {Friesen}}, \bibinfo {author} {\bibfnamefont {M.~A.}\
  \bibnamefont {Eriksson}},  \emph {et~al.},\ }\href@noop {} {\bibfield
  {journal} {\bibinfo  {journal} {npj Quantum Information}\ }\textbf {\bibinfo
  {volume} {4}},\ \bibinfo {pages} {26} (\bibinfo {year} {2018})}\BibitemShut
  {NoStop}%
\bibitem [{\citenamefont {Mi}\ \emph {et~al.}(2017)\citenamefont {Mi},
  \citenamefont {P{\'e}terfalvi}, \citenamefont {Burkard},\ and\ \citenamefont
  {Petta}}]{mi2017high}%
  \BibitemOpen
  \bibfield  {author} {\bibinfo {author} {\bibfnamefont {X.}~\bibnamefont
  {Mi}}, \bibinfo {author} {\bibfnamefont {C.~G.}\ \bibnamefont
  {P{\'e}terfalvi}}, \bibinfo {author} {\bibfnamefont {G.}~\bibnamefont
  {Burkard}}, \ and\ \bibinfo {author} {\bibfnamefont {J.~R.}\ \bibnamefont
  {Petta}},\ }\href@noop {} {\bibfield  {journal} {\bibinfo  {journal}
  {Physical review letters}\ }\textbf {\bibinfo {volume} {119}},\ \bibinfo
  {pages} {176803} (\bibinfo {year} {2017})}\BibitemShut {NoStop}%
\bibitem [{\citenamefont {Borjans}\ \emph {et~al.}(2019)\citenamefont
  {Borjans}, \citenamefont {Zajac}, \citenamefont {Hazard},\ and\ \citenamefont
  {Petta}}]{borjans2019single}%
  \BibitemOpen
  \bibfield  {author} {\bibinfo {author} {\bibfnamefont {F.}~\bibnamefont
  {Borjans}}, \bibinfo {author} {\bibfnamefont {D.}~\bibnamefont {Zajac}},
  \bibinfo {author} {\bibfnamefont {T.}~\bibnamefont {Hazard}}, \ and\ \bibinfo
  {author} {\bibfnamefont {J.~R.}\ \bibnamefont {Petta}},\ }\href@noop {}
  {\bibfield  {journal} {\bibinfo  {journal} {Physical Review Applied}\
  }\textbf {\bibinfo {volume} {11}},\ \bibinfo {pages} {044063} (\bibinfo
  {year} {2019})}\BibitemShut {NoStop}%
\bibitem [{\citenamefont {Mi}\ \emph {et~al.}(2018)\citenamefont {Mi},
  \citenamefont {Kohler},\ and\ \citenamefont {Petta}}]{mi2018landau}%
  \BibitemOpen
  \bibfield  {author} {\bibinfo {author} {\bibfnamefont {X.}~\bibnamefont
  {Mi}}, \bibinfo {author} {\bibfnamefont {S.}~\bibnamefont {Kohler}}, \ and\
  \bibinfo {author} {\bibfnamefont {J.~R.}\ \bibnamefont {Petta}},\ }\href@noop
  {} {\bibfield  {journal} {\bibinfo  {journal} {Physical Review B}\ }\textbf
  {\bibinfo {volume} {98}},\ \bibinfo {pages} {161404} (\bibinfo {year}
  {2018})}\BibitemShut {NoStop}%
\bibitem [{\citenamefont {Friesen}\ \emph {et~al.}(2007)\citenamefont
  {Friesen}, \citenamefont {Chutia}, \citenamefont {Tahan},\ and\ \citenamefont
  {Coppersmith}}]{Friesen2007ValleyWells}%
  \BibitemOpen
  \bibfield  {author} {\bibinfo {author} {\bibfnamefont {M.}~\bibnamefont
  {Friesen}}, \bibinfo {author} {\bibfnamefont {S.}~\bibnamefont {Chutia}},
  \bibinfo {author} {\bibfnamefont {C.}~\bibnamefont {Tahan}}, \ and\ \bibinfo
  {author} {\bibfnamefont {S.~N.}\ \bibnamefont {Coppersmith}},\ }\href
  {\doibase 10.1103/PhysRevB.75.115318} {\bibfield  {journal} {\bibinfo
  {journal} {Physical Review B}\ }\textbf {\bibinfo {volume} {75}},\ \bibinfo
  {pages} {115318} (\bibinfo {year} {2007})}\BibitemShut {NoStop}%
\bibitem [{\citenamefont {Weitz}\ \emph {et~al.}(1996)\citenamefont {Weitz},
  \citenamefont {Haug}, \citenamefont {von. Klitzing},\ and\ \citenamefont
  {Schäffler}}]{weitz1996tilted}%
  \BibitemOpen
  \bibfield  {author} {\bibinfo {author} {\bibfnamefont {P.}~\bibnamefont
  {Weitz}}, \bibinfo {author} {\bibfnamefont {R.~J.}\ \bibnamefont {Haug}},
  \bibinfo {author} {\bibfnamefont {K.}~\bibnamefont {von. Klitzing}}, \ and\
  \bibinfo {author} {\bibfnamefont {F.}~\bibnamefont {Schäffler}},\ }\href
  {\doibase 10.1016/0039-6028(96)00465-7} {\bibfield  {journal} {\bibinfo
  {journal} {Surface Science}\ }\textbf {\bibinfo {volume} {361-362}},\
  \bibinfo {pages} {542} (\bibinfo {year} {1996})}\BibitemShut {NoStop}%
\bibitem [{\citenamefont {Lai}\ \emph {et~al.}(2004)\citenamefont {Lai},
  \citenamefont {Pan}, \citenamefont {Tsui}, \citenamefont {Lyon},
  \citenamefont {Mühlberger},\ and\ \citenamefont {Schäffler}}]{lai2004two}%
  \BibitemOpen
  \bibfield  {author} {\bibinfo {author} {\bibfnamefont {K.}~\bibnamefont
  {Lai}}, \bibinfo {author} {\bibfnamefont {W.}~\bibnamefont {Pan}}, \bibinfo
  {author} {\bibfnamefont {D.~C.}\ \bibnamefont {Tsui}}, \bibinfo {author}
  {\bibfnamefont {S.}~\bibnamefont {Lyon}}, \bibinfo {author} {\bibfnamefont
  {M.}~\bibnamefont {Mühlberger}}, \ and\ \bibinfo {author} {\bibfnamefont
  {F.}~\bibnamefont {Schäffler}},\ }\href {\doibase
  10.1103/PhysRevLett.93.156805} {\bibfield  {journal} {\bibinfo  {journal}
  {Physical Review Letters}\ }\textbf {\bibinfo {volume} {93}},\ \bibinfo
  {pages} {156805} (\bibinfo {year} {2004})}\BibitemShut {NoStop}%
\bibitem [{\citenamefont {Sasaki}\ \emph {et~al.}(2009)\citenamefont {Sasaki},
  \citenamefont {Masutomi}, \citenamefont {Toyama}, \citenamefont {Sawano},
  \citenamefont {Shiraki},\ and\ \citenamefont
  {Okamoto}}]{Sasaki2009Well-widthWells}%
  \BibitemOpen
  \bibfield  {author} {\bibinfo {author} {\bibfnamefont {K.}~\bibnamefont
  {Sasaki}}, \bibinfo {author} {\bibfnamefont {R.}~\bibnamefont {Masutomi}},
  \bibinfo {author} {\bibfnamefont {K.}~\bibnamefont {Toyama}}, \bibinfo
  {author} {\bibfnamefont {K.}~\bibnamefont {Sawano}}, \bibinfo {author}
  {\bibfnamefont {Y.}~\bibnamefont {Shiraki}}, \ and\ \bibinfo {author}
  {\bibfnamefont {T.}~\bibnamefont {Okamoto}},\ }\href {\doibase
  10.1063/1.3270539} {\bibfield  {journal} {\bibinfo  {journal} {Applied
  Physics Letters}\ }\textbf {\bibinfo {volume} {95}},\ \bibinfo {pages}
  {222109} (\bibinfo {year} {2009})}\BibitemShut {NoStop}%
\bibitem [{\citenamefont {Neyens}\ \emph {et~al.}(2018)\citenamefont {Neyens},
  \citenamefont {Foote}, \citenamefont {Thorgrimsson}, \citenamefont {Knapp},
  \citenamefont {McJunkin}, \citenamefont {Vandersypen}, \citenamefont {Amin},
  \citenamefont {Thomas}, \citenamefont {Clarke}, \citenamefont {Savage},
  \citenamefont {Lagally}, \citenamefont {Friesen}, \citenamefont
  {Coppersmith},\ and\ \citenamefont {Eriksson}}]{Neyens2018TheWells}%
  \BibitemOpen
  \bibfield  {author} {\bibinfo {author} {\bibfnamefont {S.~F.}\ \bibnamefont
  {Neyens}}, \bibinfo {author} {\bibfnamefont {R.~H.}\ \bibnamefont {Foote}},
  \bibinfo {author} {\bibfnamefont {B.}~\bibnamefont {Thorgrimsson}}, \bibinfo
  {author} {\bibfnamefont {T.~J.}\ \bibnamefont {Knapp}}, \bibinfo {author}
  {\bibfnamefont {T.}~\bibnamefont {McJunkin}}, \bibinfo {author}
  {\bibfnamefont {L.~M.~K.}\ \bibnamefont {Vandersypen}}, \bibinfo {author}
  {\bibfnamefont {P.}~\bibnamefont {Amin}}, \bibinfo {author} {\bibfnamefont
  {N.~K.}\ \bibnamefont {Thomas}}, \bibinfo {author} {\bibfnamefont {J.~S.}\
  \bibnamefont {Clarke}}, \bibinfo {author} {\bibfnamefont {D.~E.}\
  \bibnamefont {Savage}}, \bibinfo {author} {\bibfnamefont {M.~G.}\
  \bibnamefont {Lagally}}, \bibinfo {author} {\bibfnamefont {M.}~\bibnamefont
  {Friesen}}, \bibinfo {author} {\bibfnamefont {S.~N.}\ \bibnamefont
  {Coppersmith}}, \ and\ \bibinfo {author} {\bibfnamefont {M.~A.}\ \bibnamefont
  {Eriksson}},\ }\href {\doibase 10.1063/1.5033447} {\bibfield  {journal}
  {\bibinfo  {journal} {Applied Physics Letters}\ }\textbf {\bibinfo {volume}
  {112}},\ \bibinfo {pages} {243107} (\bibinfo {year} {2018})}\BibitemShut
  {NoStop}%
\bibitem [{\citenamefont {Paquelet~Wuetz}\ \emph {et~al.}(2020)\citenamefont
  {Paquelet~Wuetz}, \citenamefont {Bavdaz}, \citenamefont {Yeoh}, \citenamefont
  {Schouten}, \citenamefont {van~der Does}, \citenamefont {Tiggelman},
  \citenamefont {Sabbagh}, \citenamefont {Sammak}, \citenamefont {Almudever},
  \citenamefont {Sebastiano}, \citenamefont {Clarke}, \citenamefont
  {Veldhorst},\ and\ \citenamefont {Scappucci}}]{wuetz2019multiplexed}%
  \BibitemOpen
  \bibfield  {author} {\bibinfo {author} {\bibfnamefont {B.}~\bibnamefont
  {Paquelet~Wuetz}}, \bibinfo {author} {\bibfnamefont {P.~L.}\ \bibnamefont
  {Bavdaz}}, \bibinfo {author} {\bibfnamefont {L.~A.}\ \bibnamefont {Yeoh}},
  \bibinfo {author} {\bibfnamefont {R.}~\bibnamefont {Schouten}}, \bibinfo
  {author} {\bibfnamefont {H.}~\bibnamefont {van~der Does}}, \bibinfo {author}
  {\bibfnamefont {M.}~\bibnamefont {Tiggelman}}, \bibinfo {author}
  {\bibfnamefont {D.}~\bibnamefont {Sabbagh}}, \bibinfo {author} {\bibfnamefont
  {A.}~\bibnamefont {Sammak}}, \bibinfo {author} {\bibfnamefont {C.~G.}\
  \bibnamefont {Almudever}}, \bibinfo {author} {\bibfnamefont {F.}~\bibnamefont
  {Sebastiano}}, \bibinfo {author} {\bibfnamefont {J.~S.}\ \bibnamefont
  {Clarke}}, \bibinfo {author} {\bibfnamefont {M.}~\bibnamefont {Veldhorst}}, \
  and\ \bibinfo {author} {\bibfnamefont {G.}~\bibnamefont {Scappucci}},\ }\href
  {\doibase 10.1038/s41534-020-0274-4} {\bibfield  {journal} {\bibinfo
  {journal} {npj Quantum Information}\ }\textbf {\bibinfo {volume} {6}},\
  \bibinfo {pages} {43} (\bibinfo {year} {2020})}\BibitemShut {NoStop}%
\bibitem [{\citenamefont {Sammak}\ \emph {et~al.}(2019)\citenamefont {Sammak},
  \citenamefont {Sabbagh}, \citenamefont {Hendrickx}, \citenamefont {Lodari},
  \citenamefont {Wuetz}, \citenamefont {Tosato}, \citenamefont {Yeoh},
  \citenamefont {Bollani}, \citenamefont {Virgilio}, \citenamefont {Schubert},
  \citenamefont {Zaumseil}, \citenamefont {Capellini}, \citenamefont
  {Veldhorst},\ and\ \citenamefont {Scappucci}}]{Sammak2019ShallowTechnology}%
  \BibitemOpen
  \bibfield  {author} {\bibinfo {author} {\bibfnamefont {A.}~\bibnamefont
  {Sammak}}, \bibinfo {author} {\bibfnamefont {D.}~\bibnamefont {Sabbagh}},
  \bibinfo {author} {\bibfnamefont {N.~W.}\ \bibnamefont {Hendrickx}}, \bibinfo
  {author} {\bibfnamefont {M.}~\bibnamefont {Lodari}}, \bibinfo {author}
  {\bibfnamefont {B.~P.}\ \bibnamefont {Wuetz}}, \bibinfo {author}
  {\bibfnamefont {A.}~\bibnamefont {Tosato}}, \bibinfo {author} {\bibfnamefont
  {L.}~\bibnamefont {Yeoh}}, \bibinfo {author} {\bibfnamefont {M.}~\bibnamefont
  {Bollani}}, \bibinfo {author} {\bibfnamefont {M.}~\bibnamefont {Virgilio}},
  \bibinfo {author} {\bibfnamefont {M.~A.}\ \bibnamefont {Schubert}}, \bibinfo
  {author} {\bibfnamefont {P.}~\bibnamefont {Zaumseil}}, \bibinfo {author}
  {\bibfnamefont {G.}~\bibnamefont {Capellini}}, \bibinfo {author}
  {\bibfnamefont {M.}~\bibnamefont {Veldhorst}}, \ and\ \bibinfo {author}
  {\bibfnamefont {G.}~\bibnamefont {Scappucci}},\ }\href {\doibase
  10.1002/adfm.201807613} {\bibfield  {journal} {\bibinfo  {journal} {Advanced
  Functional Materials}\ }\textbf {\bibinfo {volume} {29}},\ \bibinfo {pages}
  {1807613} (\bibinfo {year} {2019})}\BibitemShut {NoStop}%
\bibitem [{Note1()}]{Note1}%
  \BibitemOpen
  \bibinfo {note} {See Supplemental Material [url] for the analysis of the
  HAADF-STEM intensity profile along the heterostructure growth
  direction}\BibitemShut {NoStop}%
\bibitem [{\citenamefont {Tracy}\ \emph {et~al.}(2009)\citenamefont {Tracy},
  \citenamefont {Hwang}, \citenamefont {Eng}, \citenamefont {Ten~Eyck},
  \citenamefont {Nordberg}, \citenamefont {Childs}, \citenamefont {Carroll},
  \citenamefont {Lilly},\ and\ \citenamefont
  {Das~Sarma}}]{Tracy2009ObservationMOSFET}%
  \BibitemOpen
  \bibfield  {author} {\bibinfo {author} {\bibfnamefont {L.~A.}\ \bibnamefont
  {Tracy}}, \bibinfo {author} {\bibfnamefont {E.~H.}\ \bibnamefont {Hwang}},
  \bibinfo {author} {\bibfnamefont {K.}~\bibnamefont {Eng}}, \bibinfo {author}
  {\bibfnamefont {G.~A.}\ \bibnamefont {Ten~Eyck}}, \bibinfo {author}
  {\bibfnamefont {E.~P.}\ \bibnamefont {Nordberg}}, \bibinfo {author}
  {\bibfnamefont {K.}~\bibnamefont {Childs}}, \bibinfo {author} {\bibfnamefont
  {M.~S.}\ \bibnamefont {Carroll}}, \bibinfo {author} {\bibfnamefont {M.~P.}\
  \bibnamefont {Lilly}}, \ and\ \bibinfo {author} {\bibfnamefont
  {S.}~\bibnamefont {Das~Sarma}},\ }\href {\doibase 10.1103/PhysRevB.79.235307}
  {\bibfield  {journal} {\bibinfo  {journal} {Physical Review B}\ }\textbf
  {\bibinfo {volume} {79}},\ \bibinfo {pages} {235307} (\bibinfo {year}
  {2009})}\BibitemShut {NoStop}%
\bibitem [{Note2()}]{Note2}%
  \BibitemOpen
  \bibinfo {note} {See Supplemental Material [URL] for theoretical
  justification of this fitting form}\BibitemShut {NoStop}%
\bibitem [{Note3()}]{Note3}%
  \BibitemOpen
  \bibinfo {note} {See Supplemental Material [URL] for $g$-factor
  analysis}\BibitemShut {NoStop}%
\bibitem [{\citenamefont {Goswami}\ \emph {et~al.}(2007)\citenamefont
  {Goswami}, \citenamefont {Slinker}, \citenamefont {Friesen}, \citenamefont
  {{McGuire}}, \citenamefont {Truitt}, \citenamefont {Tahan}, \citenamefont
  {Klein}, \citenamefont {Chu}, \citenamefont {Mooney}, \citenamefont {{van der
  Weide}}, \citenamefont {Joynt}, \citenamefont {Coppersmith},\ and\
  \citenamefont {Eriksson}}]{goswami_2007}%
  \BibitemOpen
  \bibfield  {author} {\bibinfo {author} {\bibfnamefont {S.}~\bibnamefont
  {Goswami}}, \bibinfo {author} {\bibfnamefont {K.~A.}\ \bibnamefont
  {Slinker}}, \bibinfo {author} {\bibfnamefont {M.}~\bibnamefont {Friesen}},
  \bibinfo {author} {\bibfnamefont {L.~M.}\ \bibnamefont {{McGuire}}}, \bibinfo
  {author} {\bibfnamefont {J.~L.}\ \bibnamefont {Truitt}}, \bibinfo {author}
  {\bibfnamefont {C.}~\bibnamefont {Tahan}}, \bibinfo {author} {\bibfnamefont
  {L.~J.}\ \bibnamefont {Klein}}, \bibinfo {author} {\bibfnamefont {J.~O.}\
  \bibnamefont {Chu}}, \bibinfo {author} {\bibfnamefont {P.~M.}\ \bibnamefont
  {Mooney}}, \bibinfo {author} {\bibfnamefont {D.~W.}\ \bibnamefont {{van der
  Weide}}}, \bibinfo {author} {\bibfnamefont {R.}~\bibnamefont {Joynt}},
  \bibinfo {author} {\bibfnamefont {S.~N.}\ \bibnamefont {Coppersmith}}, \ and\
  \bibinfo {author} {\bibfnamefont {M.~A.}\ \bibnamefont {Eriksson}},\ }\href
  {\doibase 10.1038/nphys475} {\bibfield  {journal} {\bibinfo  {journal}
  {Nature Physics}\ }\textbf {\bibinfo {volume} {3}},\ \bibinfo {pages} {41}
  (\bibinfo {year} {2007})}\BibitemShut {NoStop}%
\bibitem [{\citenamefont {Chklovskii}\ \emph {et~al.}(1992)\citenamefont
  {Chklovskii}, \citenamefont {Shklovskii},\ and\ \citenamefont
  {Glazman}}]{chklovskii1992electrostatics}%
  \BibitemOpen
  \bibfield  {author} {\bibinfo {author} {\bibfnamefont {D.~B.}\ \bibnamefont
  {Chklovskii}}, \bibinfo {author} {\bibfnamefont {B.~I.}\ \bibnamefont
  {Shklovskii}}, \ and\ \bibinfo {author} {\bibfnamefont {L.~I.}\ \bibnamefont
  {Glazman}},\ }\href {\doibase 10.1103/PhysRevB.46.4026} {\bibfield  {journal}
  {\bibinfo  {journal} {Physical Review B}\ }\textbf {\bibinfo {volume} {46}},\
  \bibinfo {pages} {4026} (\bibinfo {year} {1992})}\BibitemShut {NoStop}%
\bibitem [{\citenamefont {Davies}(1998)}]{davies1998physics}%
  \BibitemOpen
  \bibfield  {author} {\bibinfo {author} {\bibfnamefont {J.~H.}\ \bibnamefont
  {Davies}},\ }\href@noop {} {\emph {\bibinfo {title} {The {Physics} of
  {Low}-dimensional {Semiconductors}: {An} {Introduction}}}}\ (\bibinfo
  {publisher} {Cambridge University Press},\ \bibinfo {year}
  {1998})\BibitemShut {NoStop}%
\bibitem [{\citenamefont {Gamble}\ \emph {et~al.}(2013)\citenamefont {Gamble},
  \citenamefont {Eriksson}, \citenamefont {Coppersmith},\ and\ \citenamefont
  {Friesen}}]{gamble2013disorder}%
  \BibitemOpen
  \bibfield  {author} {\bibinfo {author} {\bibfnamefont {J.~K.}\ \bibnamefont
  {Gamble}}, \bibinfo {author} {\bibfnamefont {M.~A.}\ \bibnamefont
  {Eriksson}}, \bibinfo {author} {\bibfnamefont {S.~N.}\ \bibnamefont
  {Coppersmith}}, \ and\ \bibinfo {author} {\bibfnamefont {M.}~\bibnamefont
  {Friesen}},\ }\href {\doibase 10.1103/PhysRevB.88.035310} {\bibfield
  {journal} {\bibinfo  {journal} {Physical Review B}\ }\textbf {\bibinfo
  {volume} {88}},\ \bibinfo {pages} {035310} (\bibinfo {year}
  {2013})}\BibitemShut {NoStop}%
\bibitem [{Note4()}]{Note4}%
  \BibitemOpen
  \bibinfo {note} {See Supplemental Material [URL] for theoretical methods,
  which includes Ref.~\cite {frees2019compressed}}\BibitemShut {NoStop}%
\bibitem [{\citenamefont {Frees}\ \emph {et~al.}(2019)\citenamefont {Frees},
  \citenamefont {Gamble}, \citenamefont {Ward}, \citenamefont {Blume-Kohout},
  \citenamefont {Eriksson}, \citenamefont {Friesen},\ and\ \citenamefont
  {Coppersmith}}]{frees2019compressed}%
  \BibitemOpen
  \bibfield  {author} {\bibinfo {author} {\bibfnamefont {A.}~\bibnamefont
  {Frees}}, \bibinfo {author} {\bibfnamefont {J.~K.}\ \bibnamefont {Gamble}},
  \bibinfo {author} {\bibfnamefont {D.~R.}\ \bibnamefont {Ward}}, \bibinfo
  {author} {\bibfnamefont {R.}~\bibnamefont {Blume-Kohout}}, \bibinfo {author}
  {\bibfnamefont {M.}~\bibnamefont {Eriksson}}, \bibinfo {author}
  {\bibfnamefont {M.}~\bibnamefont {Friesen}}, \ and\ \bibinfo {author}
  {\bibfnamefont {S.}~\bibnamefont {Coppersmith}},\ }\href@noop {} {\bibfield
  {journal} {\bibinfo  {journal} {Physical Review Applied}\ }\textbf {\bibinfo
  {volume} {11}},\ \bibinfo {pages} {024063} (\bibinfo {year}
  {2019})}\BibitemShut {NoStop}%
\end{thebibliography}%

\end{document}